\documentclass[twocolumn]{revtex4}
\usepackage{graphicx}

\begin{document}
\title{ Creating high dimensional time-bin entanglement using mode-locked lasers }
\author{Hugues de Riedmatten, Ivan Marcikic, Hugo Zbinden, Nicolas Gisin}
\affiliation{Group of Applied Physics, University of Geneva, 1211 Geneva
4, Switzerland}
\begin{abstract}
We present a new scheme to generate high dimensional
entanglement between two photonic systems. The idea is based on
parametric down conversion with a sequence of pump pulses
generated by a mode-locked laser. We prove experimentally the
feasibility of this scheme by performing a Franson-type Bell
test using a 2-way interferometer with path-length
difference equal to the distance between 2 pump pulses. With this
experiment, we can demonstrate entanglement for a two-photon state of at least dimension
D=11. Finally, we propose a feasible experiment to show a Fabry-Perot like effect for a high dimensional two-photon
state.

\end{abstract}
\maketitle
\subsection*{Introduction}
Photonic entanglement is one of the most important tools for quantum
communication experiments \cite{tittel01}. Different types of entanglement
can be used: for instance the well known polarization entanglement, momentum
entanglement or entanglement between photons created in a coherent
superposition of emission times, the so-called time-bin entanglement \cite{brendel99}. To
date, most of the experiments involve entangled two-level systems (qubits).
However, entanglement in high-dimensional Hilbert spaces has received a lot
of attention recently. The study of such states is interesting from a
fundamental point of view, because they lead to quantum predictions that
differ from classical physics more radically than two-level-systems and are more resistant to noise\cite
{Kaszlikowski00,collins02}. It has also recently been shown that increasing the dimension
of the Hilbert space would allow a decrease in the detection efficiency
required to close the detection loophole for EPR experiments \cite{massar01}.
Moreover, high dimensional states can be used to increase the
security of quantum key distribution \cite{cerf01}. Finally, new
quantum protocols involving qutrits (D=3) have been recently proposed
\cite{fitzi01}.
One way to reach experimentally larger Hilbert
spaces is to use higher order parametric down conversion to
generate multiqubit entanglement \cite
{Lamas01,Howell02,Weinfurter01}. In this case, however, the
increase of the Hilbert space dimension is accompanied by an
increase of the number of particles involved. A more direct way is
to use multilevel systems. Entanglement of angular momentum
states of photons has been recently suggested and demonstrated in
this context \cite{Mair01,Vaziri01}.\\ Here we propose a new scheme, allowing
in principle to reach entanglement in arbitrarily high dimensional
Hilbert spaces that is based on time-bin entanglement. One of the
advantages of this type of entanglement is that it can be
straight-forwardly extended to higher dimension Hilbert spaces
by increasing the numbers of possible times of creation. In our
previous experiments, time-bin entangled qubits were created by
inserting a two-way unbalanced interferometer between the pulsed
laser and the down-converter \cite{brendel99,tittel00}. Photons
were then sent to analyzers, which were equally unbalanced 2-way
interferometers. By inserting a D-way interferometer before the
downconverter, we could in principle obtain time-bin entangled
quDits. In this paper, we present a natural extension of
this scheme, in which the possible times of creation of the photon
pairs are directly given by the successive pump laser pulses. This
allows the creation of quantum states in arbitrarily high
dimension Hilbert spaces. A necessary condition is that the different pulses must be
coherent with each others, i.e exhibit a constant phase shift.
This coherence can be obtained by using a mode-locked pulsed
laser. To prove the feasibility of this scheme, we make a first
experiment analyzing the created high-dimensional entangled states
with a 2-way interferometer, with path-length difference equal to the distance
between two laser pulses. The paper is organized as follow: first we
explain the principle of high-dimensional time-bin entanglement;
then, we describe the experiment and finally we propose a feasible experiment to show
 Fabry-Perot like behavior for a high-dimensional two-photon state.

\subsection*{Principle of high-dimensional time-bin entanglement generated by
a sequence of coherent pulses} 
\begin{figure}[h]
\begin{center}
\includegraphics[width=8.43cm]{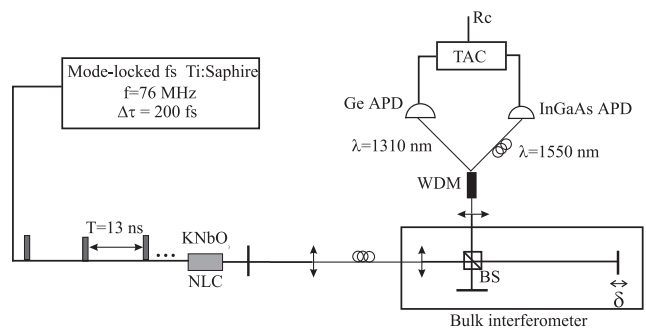}
\caption{Experimental setup }
\label{setup}
\end{center}
\end{figure}

We present here an intuitive
approach of the principle of high dimensional time-bin entanglement generated by a
sequence of laser pulses. A parametric down converter
(PDC) is pumped by a sequence of pump laser pulses with a fixed
phase difference. This sequence can be generated by a mode-locked
laser, or, for instance by a single-shot laser followed by a loop
made of a beam splitter where one output is connected to one
input \cite{comment1}. Let us assume that we have a train of D pump pulses.
Eventually a pump photon creates a photon pair by
spontaneous parametric down conversion. Let us label the
successive laser pulses by N=1,2,...,j,...D with N=1 at t=0. The
time difference between two pulses is $\Delta t.$ With this
notation, the pulse j arrives before the pulse j+1 at the
down-converter. We suppose that the probability of creating
more than one pair in D pulses is negligible. If a photon pair is created in pulse, or time-bin, j we write the state:
\begin{eqnarray}
a^{\dagger}_{j}b^{\dagger}_{j}\left|0_{a},0_{b}\right\rangle=\left|0_{1},0_{2},...,1_{j},..,0_{D}\right\rangle_{a}\nonumber
\\
\otimes
\left|0_{1},0_{2},...,1_{j},..,0_{D}\right\rangle_{b}\equiv\left|j_{a},j_{b}\right\rangle\equiv\left|j,j\right\rangle\
\end{eqnarray}
where $a^{\dagger}_{j}$ and $b^{\dagger}_{j}$ are the creation
operators for the two PDC modes in the time-bin j (in our case,
the two modes corresponds to different wavelengths, 1310nm and
1550 nm). In the
following, we will use the notation $\left|j,j\right\rangle$ for a
photon pair created in time-bin j. As we cannot know in which time
bin the photon pair is created, we have after the PDC the D-dimensional entangled state:
\begin{equation}
\left| \Psi_{PDC} \right\rangle =\sum_{j=1}^{D}c_{j}e^{i\phi_{j}}\left|
j,j\right\rangle \label{state1}
\end{equation}
where $\phi_{j}$ is the phase difference between two pump pulses and $\sum_{j=1}^{D}c_{j}^{2}=1$. It must be
stressed that this scheme enables in principle the creation of any
D-dimensional state of the form (\ref{state1}). Indeed, by
inserting an amplitude and/or phase modulator before the
down-converter, we could control the desired number of pump pulses
(and thus choose the dimension D) and modulate their amplitude and
phase (and thus setting the coefficients $c_{j}$ and $\phi_{j}$ in order to
generate non-maximally entangled states). In our experiment however, we will consider
$c_{j}$ and $\phi_{j}$ as constant.
\\ The creation of high-dimensional time-bin entangled states is relatively straight-forward.
A more difficult task is to analyze the created states in order to show high-dimensional entanglement.
The more natural way to do it would be to use D-way interferometers. However, building such
devices would lead to severe practical difficulties as D increases. Consequently, we decided to use a two-way interferometer
as analyzer. Nevertheless, it turns out that this experimental configuration allows us to demonstrate
high-dimensional entanglement, as we will explain in the following. The long arm of the two-way interferometer we used
 introduces a delay $\Delta t$ with respect to the short one (see fig. \ref{setup}).

 This means that a photon traveling through
 the short arm will remain in the same time-bin while a photon traveling
 through the long arm will move to the next time-bin.
 For a photon pair in the time-bin j, the effect of the interferometer can be written as
follow :
\begin{eqnarray}
\left| j,j\right\rangle\rightarrow\left| j,j\right\rangle
+e^{2i\delta}\left | j+1,j+1\right\rangle\nonumber\\
+e^{i\delta}\left|j,j+1\right\rangle +e^{i\delta}\left|
j+1,j\right\rangle
\end{eqnarray}
where $\delta$ is the phase shift acquired in the long arm of the
interferometer. If two photons are in the same time-bin after the
interferometer, they will be detected with a time difference $\tau
=0$, while if they are in different time bins, they will be
detected with a time difference $\tau=\Delta t$. In the following,
we will consider only the terms leading to a coincidence
with $\tau = 0$, because only those terms lead to
indistinguishable processes. In practice, this is achieved by
selecting a sufficiently small coincidence window. After the
interferometer, a state of the form (\ref{state1}) becomes:
\begin{eqnarray}
\left|\Psi_{PDC}\right\rangle\rightarrow \left|\Psi_{int}\right\rangle
=\left|1,1\right\rangle+\sum_{j=2}^{D}\left|
j,j\right\rangle(1+e^{2i\delta})\nonumber\\
+e^{2i\delta}\left|D+1,D+1\right\rangle
\label{psiint}
\end{eqnarray}
We see that for all time-bins except the first and the last one we
have a superposition of two indistinguishable processes. In
principle, the first and the last time bin (i.e photons created in
the first pulse traveling through the short arm and photons created
in the last pulses traveling through the long arm) can be discarded
by using switches. In this case, the coincidence count rate is
given by:
\begin{equation}
R_{c}\sim 1+V\cos (2\delta)
\end{equation}
where V is the visibility of the interference pattern given by known experimental imperfections.
The maximum
visibility in this case is 100$\%$. In practice, if we don't
discard the first and the last time-bins, the maximal visibility
will be reduced. We can see from eq. \ref{psiint} that we have 2D different
processes that lead to a coincidence with $\tau=0$. Among these 2D processes,
there are always two (the first and the last time-bins) which are
completely distinguishable. Hence the maximum visibility is
given by:
\begin{equation}
V=\frac{D-1}{D}
\label{visD}
\end{equation}
Consequently, by measuring a given
visibility in this experiment, we can prove that \emph{we generate entanglement for a two-photon state of at least dimension
 $D=\frac{1}{1-V}$.} In our experiment however, we don't
have trains of exactly D pulses. Actually, as we pump the PDC directly with a pulsed laser, we have an (almost) infinite
number of pulses. Here the dimension D is bounded by the probability of
creating one photon pair in D pulses and by the stability of the pump laser cavity. \\
Another interesting point with this experimental configuration is
that the number of possible outcomes is higher than the dimension
of the space. We have indeed D+1 time-bins, and for each time-bin
we have two possible outcomes (a given photon can take one or the other output of the interferometer).
We have thus $2(D+1)>D$ possible
results. This constitutes a generalized quantum
measurement, a so-called positive operator-valued measure (POVM) \cite{peres}.
\subsection*{Experiment}

\begin{figure}[h]
\begin{center}

\includegraphics[width=8.43cm]{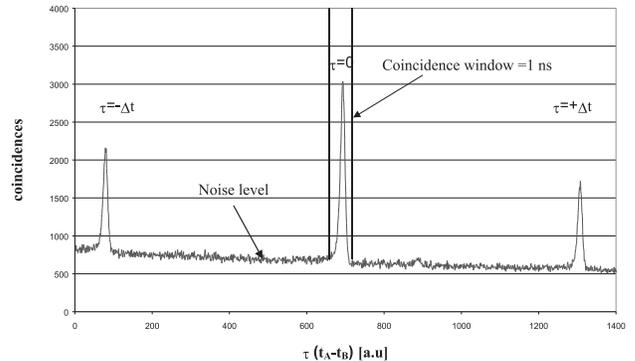}
\caption{Histogram of the two photons arrival time difference, as recorded
with the TAC. The coincidence window is centered around the $\protect\tau =0$
peak. The time between two side peaks is given by the period of the laser
(t=13 ns). The decreasing of the noise rate is due to afterpulses effects
\cite{stucki01}}
\label{TAC}
\end{center}
\end{figure}

\begin{figure}[h]
\begin{center}
\includegraphics[width=8.43cm]{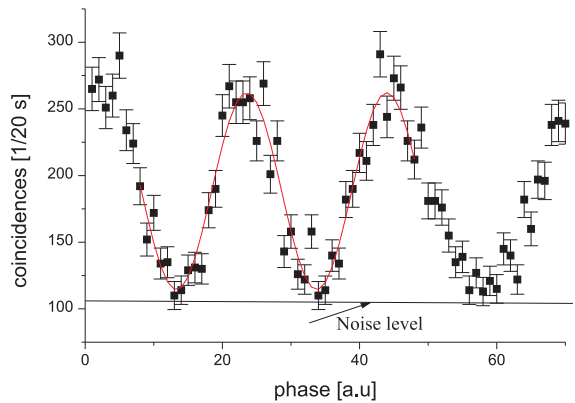} .
\caption{Coincidence count rate as a function of the phase of the
interferometer. Black squares are experimental points. The error bars are
the statistical errors. The solid line is a sinusoidal fit, from which we
can infer a visibility of 91 $\pm $ 6 \%}
\label{interf}
\end{center}
\end{figure}

The experimental setup is shown in figure \ref{setup}. A
Ti-Sapphire femtosecond laser(Mira pumped by a 8W Verdi) at $\lambda_{p}=$710 nm is used to
pump a KNbO$_{3}$ non linear crystal with type-one phase matching.
The laser pulses are separated by 13 ns and their duration is
approximately 200 fs. The crystal is cut in order to emit
non-degenerate photon pairs at telecom wavelength (1310/1550 nm).
After the crystal, a filter RG1000 is used to stop the pump beam.
The down converted photons are focused into a single mode optical
fiber and directed towards a bulk Michelson interferometer
consisting of a 50/50 beam splitter (BS) and of 2 mirrors M$_{1}$
and M$_{2}$. The path-length difference between the two arms
corresponds to half the period of the laser, i.e 6.55 ns $\longleftrightarrow $
1.95 m. It thus introduces a suitable delay so that photons produced by
 two successive laser pulses interfere at the beam splitter
 \cite{comment}.
The phase in one arm of the interferometer can be varied with a
piezzoelectric crystal controlling the position of the mirror. The
losses in the interferometer are 12 dB for each wavelength.
The output of the interferometer is focused into a single mode optical fiber and
the photon are separated deterministically by a wavelength division multiplexer (WDM).
The 1310 nm photon is detected with
liquid nitrogen cooled Ge avalanche photodiodes (APD) operating in
photon counting mode, i.e. reverse biased above the breakdown
voltage (the so-called Geiger mode) \cite{ribordy98}. The quantum efficiency of
such detectors is about 10\% for 30 kHz dark counts. Ge APD's are
not sensitive at 1550 nm, thus InGaAs APD's are used to detect the
1550 nm photons. In order to remain at a reasonably low dark count
level, this kind of APD must be operated in the so called gated
mode, i.e active only during a short time window when a photon is
expected. We use Peltier cooled Epitaxx APD's, with around 20\% quantum efficiency,
for a noise probability of $5*10^{-5}$ per nanosecond \cite{stucki01}.
In our case, the InGaAs APD is activated by a clic on
the Ge APD. Therefore, a suitable optical delay must be inserted
before the InGaAs APD, in order to detect the two photons in
coincidence. The signals of the two APD's are finally sent to a
Time-to-Amplitude Converter (TAC), in order to record the time
histogram of the arrival times of the photons. A coincidence
window of about 1 ns is selected with a single channel analyzer
(SCA) around the desired peak. The coincidence rate is finally
recorded as a function of the phase shift $\delta$ in the
interferometer. Figure \ref{TAC} shows the histogram of the two
photon's arrival time
difference, as recorded with the TAC. The central peak corresponds
to photons detected in the same time-bin
($\tau =0$) while the satellite peaks correspond to photons detected
 in different time-bins. The coincidence
window is centered around the interfering peak $\tau=0.$ Figure \ref{interf}
shows the interference curve, i.e the coincidence rate as a function of the
phase shift in the interferometer, for an average pump power of 8mW.
The pump power is kept very low, in order to avoid the creation of
multiphotons states that would reduce the visibility
\cite{marcikic02}.

We observe the predicted sinusoidal behaviour, and a sinusoidal fit gives a
net visibility \ of 91 $\pm 6$\%. The rather high level of noise in the
coincidence count rates is due to the dark counts in the detectors.
From this measured visibility, we can infer, using eq.\ref{visD},
that \emph{we generated entanglement for a two-photon state of at least dimension D=11.}

\subsection*{Two photon Fabry-Perot interferometer}
\begin{figure}[h]
\begin{center}

\includegraphics[width=8.43cm]{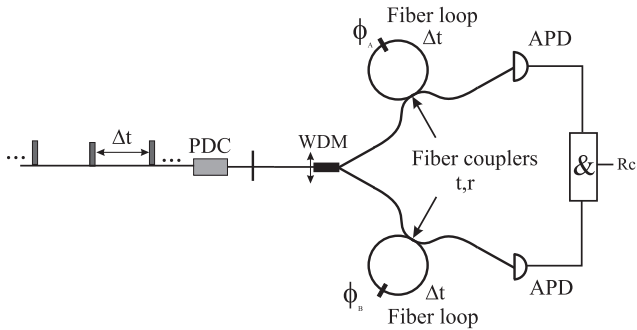}
\caption{Setup proposed to show two-photon Fabry-Perot like effects.}
\label{loop}
\end{center}
\end{figure}

High-dimensional time-bin entanglement can lead to interesting effects.
As an example, we propose a feasible experiment that shows a Fabry-Perot like
effect for two photon states. This effect is a direct manifestation of high-dimensional entanglement.
The idea is to use fiber loops as analyzers, as shown in
fig \ref{loop}.
In this case, down converted photons are separated directly after the down
converter and each one is sent to a fiber loop introducing an optical delay $%
\Delta t$ equal to the distance between pump pulses, before going to the
detector. This fiber loop is made of a fiber coupler where one output is
connected to one input. The coupler has a transmission amplitude t and a
reflection amplitude r with $t^{2}+r^{2}=1.$ By convention, a photon
transmitted remains in the same fiber. Similarly to the 2-way interferometer,
 we detect only photons in the same time-bin, i.e with $\tau =\tau_{A}-\tau_{B}=0$
 (photons having traveled the same path length from the PDC to the
detectors). The photons passing in the loop aquire a phase $\phi _{A}$ and $%
\phi _{B}$ where the phase is defined as the difference of optical delay
between 2 pump pulses and the loop. The probability of coincidences between
Alice and Bob is given by :
\begin{eqnarray}
&&P_{coinc}  \nonumber \\
&=&\left| t^{2}+r^{2}e^{i(\phi _{A}+\phi _{B})}r^{2}+r^{2}e^{i(\phi
_{A}+\phi _{B})}t^{2}e^{i(\phi _{A}+\phi _{B})}r^{2}+...\right| ^{2}
\nonumber \\
&=&\left| t^{2}+r^{4}\sum\limits_{n\geq 0}^{{}}t^{2n}e^{i(n+1)(\phi
_{A}+\phi _{B})}\right| ^{2}  \nonumber \\
&=&\left| t^{2}+\frac{r^{4}e^{i(\phi _{A}+\phi _{B})}}{1-t^{2}e^{i(\phi
_{A}+\phi _{B})}}\right| ^{2}  \label{pcoinc}
\end{eqnarray}
\begin{figure}[h]
\begin{center}

\includegraphics[width=8.43cm]{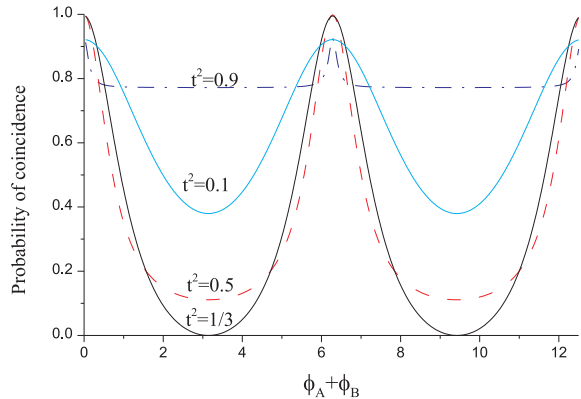}
\caption{Probability of coincidence as a function of the sum of the phases $%
\protect\phi _{A}+\protect\phi _{B}$, plotted for different transmission
probability $t^{2}$. We see that the interference has a visibility of 1 for $%
t^{2}=1/3$.}
\label{interfloop}
\end{center}
\end{figure}
Where we made the hypothesis that $t_{A}=t_{B}=t$ and $r_{A}=r_{B}=r.$
As the sum converges rapidly, the last equality is valid even with a finite number of terms $>>1$.
Figure \ref{interfloop} shows a plot of the coincidence count rate as a
function of the sum of the two phases $\phi _{A}+\phi _{B}$, computed with (%
\ref{pcoinc}). We see that the interference pattern is periodic, but no
longer sinusoidal as in the traditional case. The shape and the
visibility of the interference pattern varies strongly with the probability
of transmission in the beam splitter. The maximum visibility V =1 is
obtained for a transmission probability of $t^{2}=1/3$. When using
a two-way interferometer as analyzer, a coincidence in time-bin j is the result of the interference
of time-bins j and j-1 (nearest neighbor interference). With the fiber loop on the contrary,
 a coincidence in time-bin j is the result of the interference of all preceding time-bins.
 In this sense, a fiber loop can be
used to mimic a D way interferometer.
\subsection*{Conclusions}
We presented a new scheme to generate high-dimensional time-bin entanglement
between two photons. The method is based on parametric down conversion with
a sequence of pump pulses with a fixed phase difference. The necessary
coherence between pump pulses can be obtained by using a mode-locked laser.
This scheme enables in principle the generation of any desired two-photon high-dimensional
state by setting the number, the amplitude and the phase of the pump
pulses. We proved experimentally the feasibility of such a scheme by
analyzing the high-dimensional entangled state with a two-way interferometer.
The measured visibility of 91 $\pm 6$ \% allows us to demonstrate
entanglement for a two-photon state of at least dimension D=11. We also proposed a
feasible experiment to show a Fabry-Perot like effect for
2-photons states, as a direct consequence of high dimensional
time-bin entanglement. In conclusion, we can say that, with this
method, creating high-dimensional entanglement is relatively easy.
A more difficult and challenging task will be to use
high-dimensional entanglement in quantum protocols.
\subsection*{Aknowledgements}
Financial support by the swiss NCCR Quantum Photonics, and
by the European project QuComm is aknowledged.


\begin{thebibliography} {9}
\bibitem{comment1} In this case however, the pump intensity would
decrease constantly.
\end{thebibliography}
\end{document}